\documentclass[5p]{elsarticle}

\journal{Nuclear Instruments and Methods in Physics Research A}

\usepackage[english]{babel}
\usepackage[sc]{mathpazo}
\usepackage{amsmath}
\usepackage{units}
\usepackage[colorlinks=true]{hyperref}
\usepackage{color}
\usepackage{xspace}

\def\eq#1{\begin{equation}#1\end{equation}}

\def\vc#1{\mathbf{#1}}

\def\bmat#1{\begin{bmatrix}#1\end{bmatrix}}
\def\Re{\operatorname{Re}}

\def\Offline{\mbox{$\overline{\textrm%
{Off}}$\hspace{.05em}\protect\raisebox{.4ex}%
{$\protect\underline{\textrm{line}}$}}\xspace}
\def\geant{\textsc{Geant4}\xspace}

\begin{document}


\begin{frontmatter}

\title{Simulation of large photomultipliers for experiments in 
astroparticle physics}

\author[ung]{Alexandre Creusot}
\ead{alexandre.creusot@p-ng.si}
\author[ung,ijs]{Darko Veberi\v{c}\corref{cor}}
\ead{darko.veberic@p-ng.si}
\cortext[cor]{Corresponding author, tel.: +386 5 3315 255, fax: +386 5 
3315 385.}
\address[ung]{Laboratory for Astroparticle Physics, University of Nova 
Gorica, Slovenia}
\address[ijs]{Department of Theoretical Physics, J. Stefan Institute, 
Ljubljana, Slovenia}

\begin{abstract}
We have developed an accurate simulation model of the large 9 inch 
photomultiplier tubes (PMT) used in water-Cherenkov detectors of 
cosmic-ray induced extensive air-showers.  This work was carried out as 
part of the development of the \Offline simulation software for the 
Pierre Auger Observatory surface array, but our findings may be relevant 
also for other astrophysics experiments that employ similar large PMTs.

The implementation is realistic in terms of geometrical dimensions, 
optical processes at various surfaces, thin-film treatment of the 
photocathode, and photon reflections on the inner structure of the PMT.  
With the quantum efficiency obtained for this advanced model we have 
calibrated a much simpler and a more rudimentary model of the PMT which 
is more practical for massive simulation productions. We show that the 
quantum efficiency declared by manufactures of the PMTs is usually 
determined under conditions substantially different from those relevant 
for the particular experiment and thus requires careful 
(re)interpretation when applied to the experimental data or when used in 
simulations. In principle, the effective quantum efficiency could vary 
depending on the optical characteristics of individual events.
\end{abstract}


\end{frontmatter}

\section{Introduction}

The Pierre Auger Observatory is the largest detector built to detect
cosmic rays with energies above $\unit[10^{18}]{eV}$ \cite{ea}.  The
surface detector part consists of over 1660 water-Cherenkov detectors
distributed over an area of more than $\unit[3000]{km^2}$.  While the
longitudinal part of the shower development is measured with the
fluorescence detector, the muonic and electromagnetic lateral
components are measured at ground level with the surface detector
(SD). The individual SD stations consist of $\unit[12]{t}$ of purified
water in a light-tight container with highly reflective and diffusive
inner surface \cite{ea}. The Cherenkov light produced by the
penetrating charged particles is observed with PMTs. The signal levels
from the PMTs are constantly calibrated against the average response
to the atmospheric background muons that trigger at the lowest
threshold level \cite{vem}. Through dedicated experiments
\cite{vem,allison} the response to atmospheric background muons
arriving from all directions \cite{grieder} has been related to the
selection of atmospheric muons arriving predominantly vertically
through the center of the SD station. All signals are thus measured in
relative units of VEM (Vertical Equivalent Muon), i.e.\ relative to
the average signal that would be produced by the vertical-centered
muon. Although this auto-calibration procedure successfully removes
most of the systematics due to the detector changes, there are
nevertheless some applications that require a certain degree of
knowledge about absolute values. The absolute number of detected muons
and the size of the electromagnetic fraction in the signal are
important quantities, e.g.\ in studies of hadronic interactions in
air-shower cascades \cite{ralph} or for the purposes of primary
particle identification \cite{healy,photon}, just to name a few. We
have developed a detailed simulation of the processes of light
detection in the PMTs to reproduce known calibration data of the SD
stations and allow comparison of real and simulated events. This study
is part of the efforts to produce a complete SD simulation chain for
the Pierre Auger Observatory \cite{tripathi,ghia,billoir}.

\section{Photomultiplier tube}

The SD station has a cylindrical shape. Its height is $\unit[1.2]{m}$
and its radius is $\unit[1.8]{m}$. The Cherenkov radiation emitted in 
the water of the SD station is captured by three 9-inch PMTs floating
on top of the water container, positioned $\unit[1.2]{m}$ away from the 
cylinder axis with $120^\circ$ separation in azimuth. The PMTs have been 
produced by Photonis; the particular PMT model XP1805-PA1 used by the 
Pierre Auger Observatory differs
from their stock model XP1805 \cite{photonis-catalogue} only in the 
additional output from the last dynode. Together with the usual anode 
output these two are used for the monitoring of the dynode--anode ratio 
and the low and high gain signal acquisition \cite{genolini}.

The geometry of the PMT can be well approximated by an oblate spheroid 
with two equal semi-principal axes of $\unit[108.8]{mm}$ length and a
third axis $\unit[79.4]{mm}$ long (outer dimensions). The thickness of 
the glass window is $\unit[2.5]{mm}$. The glass is composed of 
borosilicate (80\% SiO$_2$, 13\% B$_2$O$_2$, and 7\% Na$_2$O) with an 
index of refraction of $1.47$ and a strong increase of absorption for 
wavelengths below $\sim\unit[300]{nm}$. The photocathode consists of a 
thin $\sim\unit[20]{nm}$ layer of bi-alkali metal (KCsSb) and has a 
wave\-length-dependent complex index of refraction \cite{moorhead}.

Based on this kind of geometry and material specifications of the PMT, 
we developed a simulation model that retains the basic geometry 
properties of the PMT while simulating in great detail the physical 
processes leading to the photoelectron emissions.

\section{Advanced simulation model}

\begin{figure}[t]
\centering{}
\includegraphics[width=0.48\textwidth]{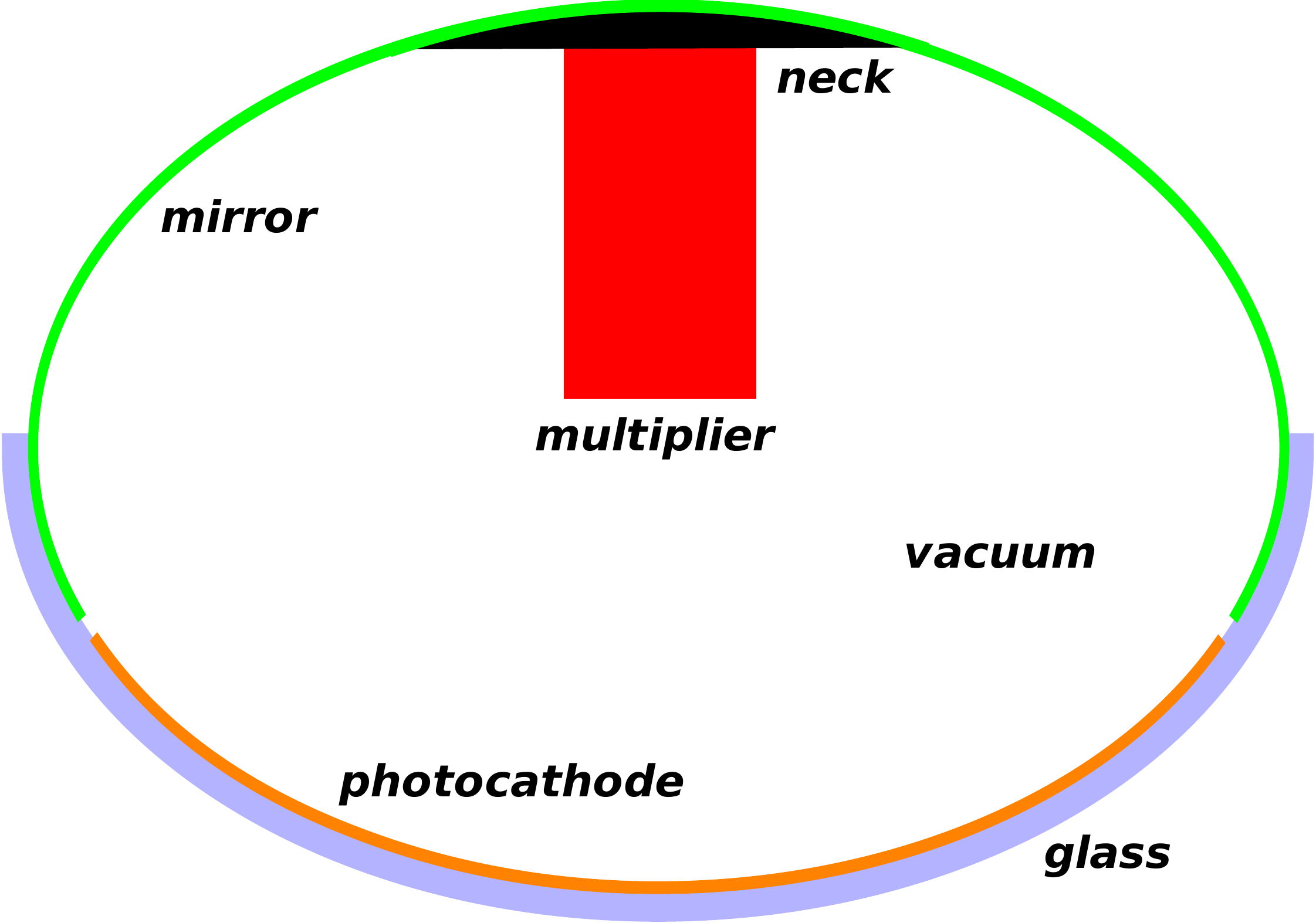}
\caption{Schema of the advanced simulation model. The main components of 
the simulation set-up are the window \emph{glass}, \emph{photocathode}, 
\emph{mirror}, effective absorption by the \emph{neck}, and the 
\emph{multiplier} structure.}
\label{f:ThinFilmPMT}
\end{figure}

The geometry set-up of the advanced simulation mod\-el is shown in 
Fig.~\ref{f:ThinFilmPMT}. The whole shape is approximated with a full 
oblate spheroid with a mirrored back wall reaching $\unit[30]{mm}$ below 
the center. The reflectivity of the \emph{mirror} is set to 97\% and 
only ideal specular reflection is implemented. As in the case of the 
real PMTs, the photocathode on the inner side of the glass is separated 
from the mirror coating by a transparent gap of $\unit[5]{mm}$.

\subsection{Inner structure of the PMT}

The loss of photons in the extended \emph{neck} leading to the PMT
base is simulated by the 100\% absorbing patch with a radius of
$\unit[40]{mm}$ placed at the top of the PMT. The \emph{multiplier}
structure (dynode stack) reaches well into the center of the PMT and
is in this particular PMT enclosed in a metallic shield case of
cylindrical shape. In the photon-tracking experiments we have found
that the proper inclusion of this multiplier volume can greatly
influence the number of reflected photons (see
Fig.~\ref{f:photonisExp} for a clear demonstration on how the parallel
beam of light gets focused into the multiplier volume). The multiplier
structure is thus simulated with a cylinder of height $\unit[60]{mm}$
and radius $\unit[16]{mm}$. It is made of a copper-like metal and its
effective reflectivity is set to 30\% (for simplicity specular
reflection only).

\subsection{Optical properties of the PMT glass}

\begin{figure}[t]
\centering
\includegraphics[width=0.48\textwidth]
{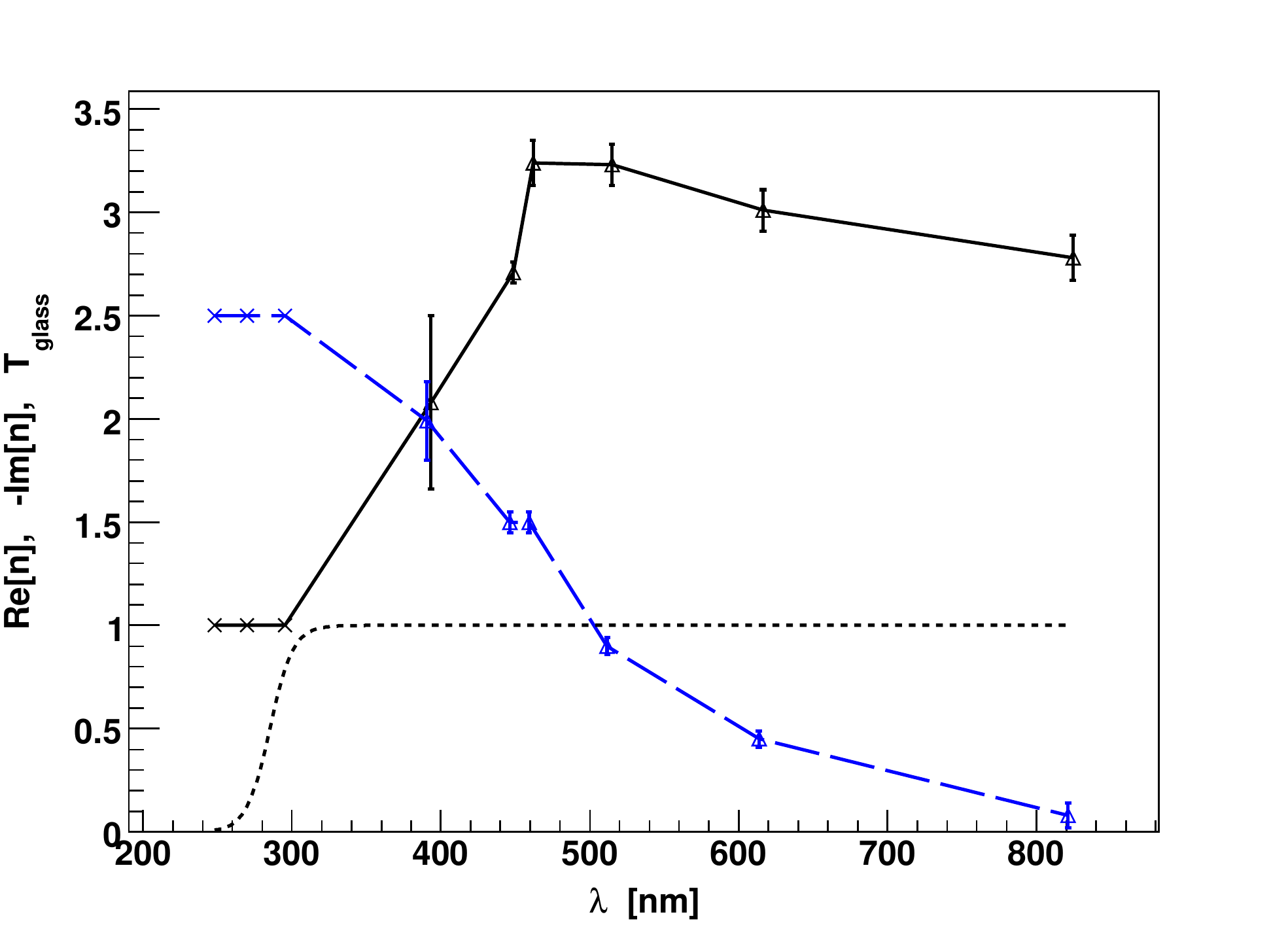}
\caption{Real (full line) and negative imaginary (dashed line) parts of 
the complex refraction index of the photocathode as a function of the 
wavelength. The triangles are experimental data points extracted from
\cite{moorhead}. The crosses are extrapolated values. The dotted line 
represents the Fermi function used to model the transmission factor of 
the PMT glass.}
\label{f:IndexesW}
\end{figure}

The material of the glass window limits the spectral sensitivity in
the short wavelength region. The borosilicate glass has a cut-off
wavelength of $\unit[270]{nm}$ (the point where it decreases below 
10\%). We have modeled this property with a simple transmission factor 
$T_\text{glass}$ that is imposed upon the entering photons,
\eq{
T_\text{glass} =
  \exp[-d/L(\lambda)] =
  \operatorname{F}(\lambda;\lambda_\text{m},\Delta\lambda),
}
where $d$ is the thickness and $L(\lambda)$ is the absorption length of
the PMT glass. The final transmission factor is modeled with a 
Fermi-function dependence on the wave\-length, 
$\operatorname{F}(\lambda;\lambda_\text{t},\Delta\lambda)= 
1/(\exp[(\lambda_\text{t}-\lambda)/\Delta\lambda]+1)$ with  
$\lambda_\text{t}=\unit[285]{nm}$ for the transition wavelength and 
$\Delta\lambda=\unit[8]{nm}$ for the transition width 
\cite{photonis-principles} (see dotted line in Fig.~\ref{f:IndexesW}).

\subsection{Photocathode as a thin film}

The sensitivity of the different types of photocathodes is restricted by 
the photo-emission threshold for the long wavelengths and can vary with 
thickness at the short wavelengths. The bi-alkali component KCsSb is 
practically the universal choice of photocathode material for Cherenkov 
light applications, and although it has been known for almost five 
decades, the availability of reliable measurements of index of 
refraction has been rather scarce. In our simulation of the angular and 
wavelength dependencies of the quantum efficiency we have used the data 
points from the experimental compilation of the complex index of 
refraction from \cite{moorhead}, where the optical properties of this 
bi-alkali material of similar thickness have been reviewed for the 
purposes of the solar neutrino detector SNO. Due to the limited number 
of different wavelength measurements we have used a simple linear 
interpolation of the complex index of refraction with wavelength 
extension along the lines of newer experimental results from 
\cite{motta} (see Fig.~\ref{f:IndexesW}).

\subsection{Photoelectron production}

\begin{figure}[t]
\centering
\includegraphics[width=0.48\textwidth]{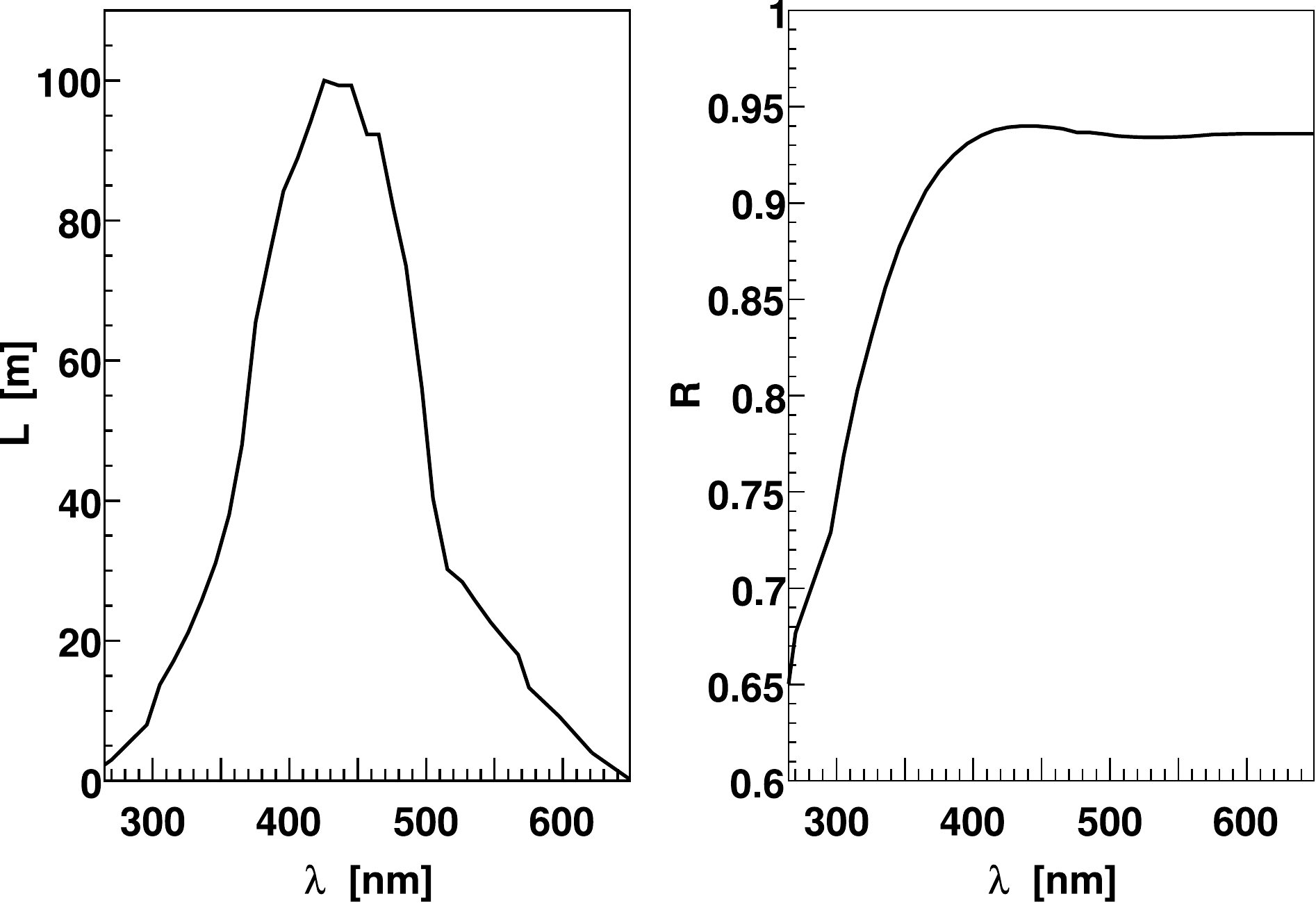}
\caption{Wavelength dependency of the water absorption length $L$ and
  the reflectivity $R$ of the container walls. Data taken from
  \cite{water} and \cite{reflectivity}, respectively.}
\label{f:abs-refl}
\end{figure}

When the photon reaches the thin layer of the photocathode it can
either get reflected with probability $R$, it can get transmitted
across the layer with effective probability $T$, or it can get
absorbed by the photocathode material with probability $A=1-R-T$. If
the photon energy in the latter case is larger than the needed exit
work, the electron can leave the photocathode. The corresponding
quantum efficiency $q_\text{pc}$ of the photocathode surface can thus
be written as
\eq{
q_\text{pc}(\lambda,\theta) = A(\lambda,\theta)\,p_\text{conv},
\label{conversion}
}
where $A(\lambda,\theta)$, the absorption coefficient, depends on
incidence angle $\theta$ and wavelength $\lambda$, and $p_\text{conv}$
is the conversion factor. The conversion factor $p_\text{conv}$ does
not depend on the angle of incidence \cite{lay,chyba} but in principle
it can still be wavelength dependent \cite{lay2}. Due to the
relatively narrow window of relevant wavelengths imposed by the water
absorption (see Fig.~\ref{f:abs-refl}) we approximate the conversion
factor $p_\text{conv}$ for the purposes of our simulations with a
constant. In section \ref{s:conversion-calibration}, we determine its
effective value from the separate \emph{simulation of the quantum
  efficiency experiment}, along the lines of the experiments usually
performed by the PMT manufacturers. In the next section we will derive
the second missing parameter $A(\lambda,\theta)$.

At this point it is worth to mention that the quantum efficiency
$q_\text{pc}$ from Eq.~\eqref{conversion} is the quantum efficiency of
the \emph{photocathode surface} and is only indirectly related to the
overall quantum efficiency of the PMT, as found in the specifications
of the manufacturers. The main difference comes from the fact that
PMTs are designed to increase light collection through reflections on
the mirror back wall. This increase is to some degree contained in the
PMT specification.  Nevertheless, its magnitude depends substantially
on the particular way the quoted efficiency is obtained
experimentally.  Therefore, the quoted efficiency specifications of
the PMT always require proper interpretation when used for a
photocathode simulation.  This fact is in general relevant also for
other physics experiments involving PMTs.

\subsection{Thin-film Fresnel equations}

Studies on PMTs usually follow \cite{born} to derive the angular and 
wavelength dependence of the optical properties. Here, we follow the 
more concise derivation from \cite{macleod} that is specifically 
dedicated to thin film treatment.  Furthermore, we adopt from 
\cite{macleod} the sign convention for the imaginary part of the index 
of refraction (see Fig.~\ref{f:IndexesW}).

Although our PMT has only one layer of thin-film photocathode, we will
for the sake of clarity derive Fresnel equations for a stack of $m$
thin-film surfaces with corresponding thicknesses $d_j$ and indexes of
reflection $n_j$. The indexes $n_j$ vary with different wavelengths
but in the expressions below we do not explicitly notate the
dependence on $\lambda$ for the sake of clarity. This stack of thin
films is on the in-going and out-going side surrounded by two
semi-infinite substrates with indexes $n_\text{in}$ and
$n_\text{out}$, respectively.

The angles of refraction in the consequent layers can all be obtained
from the incidence angle $\theta_\text{in}$ from Snell's law
\eq{
n_j\cos\theta_j = \sqrt{n_j^2-n_\text{in}^2\sin^2\theta_\text{in}}.
\label{snell}
}
This expression should be used for all $m$ layers $j=1\ldots m$ even 
when it produces a complex cosine both in case of a complex index $n_j$ 
or a total internal reflection $n_j<n_\text{in}\sin\theta_\text{in}$.  
Phase change in the layer $j$ is denoted by
\eq{
\delta_j = 2\pi\frac{d_j}{\lambda}n_j\cos\theta_j.
}
In the case of thin layers, the transmitted and reflected light combine 
coherently and the vector of normalized emergent fields 
$\vc{f}_\text{out}$ can be expressed as in \cite{macleod} with 
characteristic matrices of $m$ stacked layers as
\eq{
\vc{f}_\text{out} = \bmat{f_E\\f_H} =
\left(
\prod_j^m
\bmat{
\cos\delta_j & i\sin\delta_j/\eta_j
\\
i\eta_j\sin\delta_j & \cos\delta_j
}
\right)
\bmat{1\\\eta_\text{out}},
\label{matrix}
}
where $f_E$ and $f_H$ are normalized electrical and magnetic fields, 
respectively, and $\eta_j$, $\eta_\text{out}$ are the tilted optical 
admittances. Using short-hand notation for scalar products
\eq{
t = \vc{f}_\text{out}\cdot\bmat{\eta_\text{in}\\1}
\qquad\text{and}\qquad
r = \vc{f}_\text{out}\cdot\bmat{\eta_\text{in}\\-1} / t,
}
the reflectance is expressed as
\eq{
R = r\,r^* = |r|^2 =
  \frac{|f_E|^2\eta_\text{in}^2-2\Re[f_Ef_H^*]\eta_\text{in}+|f_H|^2}
       {|t|^2}
\label{reflectance}
}
and the transmittance as
\eq{
T = 4 \eta_\text{in}\frac{\Re[\eta_\text{out}]}{|t|^2}.
\label{transmittance}
}
Absorptance follows from
\eq{
A = 4 \eta_\text{in}\frac{\Re[f_Ef_H^* - \eta_\text{out}]}{|t|^2}.
\label{absorptance}
}
The three quantities obey the relation $R+T+A=1$. Note that while 
$\eta_\text{in}$ is always real, $\eta_j$ can be complex due to 
absorption, and $\eta_j$, $\eta_\text{out}$ can become complex when the 
incidence angle $\theta_\text{in}$ increases above the angle of total 
internal reflection $\theta_j^\text{tot}$ defined by
$\sin\theta_j^\text{tot}=n_j/n_\text{in}$.

The equations above have to be considered for two polarization cases.  
Using the convention where p-polariz\-ation implies a magnetic field 
component parallel to the interface boundary (TM wave) and 
s-polarization implies a parallel electric component (TE wave), two 
variants of the admittance emerge,
\eq{
\eta_j^\text{p} = n_j / \cos\theta_j
\qquad\text{and}\qquad
\eta_j^\text{s} = n_j\cos\theta_j,
}
where $\cos\theta_j$ is again obtained from Eq.~\eqref{snell}. The same 
polarization cases should also be used on $\eta_\text{in}$ and 
$\eta_\text{out}$. The upper expressions in \eqref{reflectance}, 
\eqref{transmittance}, and \eqref{absorptance} thus have to be 
separately evaluated for the p- and s-polarization; hence the obtained 
reflectance, transmittance, and absorptance change accordingly into 
$R_\text{p,s}$, $T_\text{p,s}$, and $A_\text{p,s}$. For unpolarized 
light we can define average quantities 
$R=\tfrac12(R_\text{p}+R_\text{s})$, 
$T=\tfrac12(T_\text{p}+T_\text{s})$, and 
$A=\tfrac12(A_\text{p}+A_\text{s})$ which, as before, satisfy $R+T+A=1$.

Cherenkov light from the injected muon is polarized perpendicularly to
the emission cone. The rotational symmetry around the muon track and
the fact that after several reflections on the inner walls the photon
polarization is after $\sim\unit[30]{ns}$ completely randomized,
enable us to use the above polarization-averaged expressions for
reflectance, transmittance, and absorptance.

\begin{figure}[t]
\centering
\includegraphics[width=0.48\textwidth]{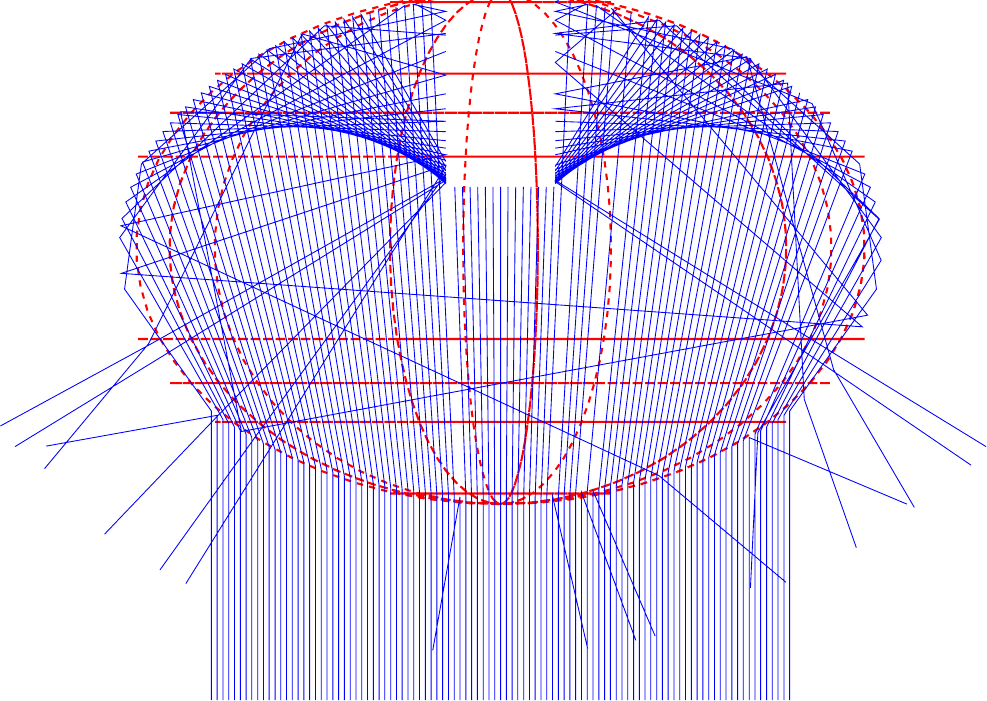}
\caption{Simulation of the Photonis quantum efficiency experiment for
  estimation of the conversion factor. Absolutely calibrated parallel
  beam of light is sent to the front of the PMT (photocathode is at
  bottom of the oval). Due to the reflection on the ellipsoidal mirror
  at the back of the PMT, the light gets approximately focused at the
  position of the multiplier (caustics at the sides of the
  multiplier). The inclusion/exclusion of the multiplier volume can
  have great effect on the amount of the light returned to the
  photocathode surface and thus has to be properly simulated. In this
  figure the multiplier (c.f.\ white patch at top center) is obscuring
  most of the returning light in this configuration of the incoming
  beam.}
\label{f:photonisExp}
\end{figure}

\subsection{Simulation set-up}

The simulation of the PMT has been implemented within the Auger
\Offline software framework \cite{argiro} where the water-Cherenkov
detector simulation is implemented with the \geant toolkit
\cite{geant4}. The production of the Cherenkov light by the injected
vertical-centered muons, the consequent absorption of the photons in
the water, and the tracking of the photons and their reflections on
the container walls have been assigned to the \geant part of the
simulation with parametrizations shown in Fig.~\ref{f:abs-refl}. The
entry of the photons into the glass of the PMT window was also handled
by \geant. From this point on, the photons were tracked by our
simulation model, based on the considerations made in the previous
sections.  Using the index of refraction parametrization from
Fig.~\ref{f:IndexesW} for $n_1$, the individual photons were reflected
from the photocathode with the probability following from the
(polarization-averaged) reflectance in Eq.~\eqref{reflectance}, or
transmitted according to Eq.~\eqref{transmittance} into the vacuum
($n_\text{out}=1$) of the PMT.  The simulation proceeded with the
photon tracking inside the PMT which can, upon successful reflections
on the inner structure, produce another crossing of the photocathode
(this time with the exchanged values for the ``in'' and ``out''
variables in the upper equations) and reentry into the water. In case
of an absorption of the photon upon the crossing of the photocathode,
the associated photoelectron is produced according to the quantum
efficiency in Eq.~\eqref{conversion}.

\subsection{Calibration of the conversion factor}
\label{s:conversion-calibration}

For the wavelength of $\unit[375]{nm}$, Photonis
\cite{photonis-catalogue} quotes a value of quantum efficiency of the
PMT as $q_\text{pmt}=0.26$.  They perform the measurement with an
absolutely-calibrated parallel beam of light with a diameter of
$\unit[170]{mm}$ illuminating the front of the PMT along the PMT axis.
Correcting for the 70\% collection efficiency (obtained from their own
electric field simulations \cite{photonis-principles}), they
consequently convert the measured current into a released
photoelectron flux $\Phi_\text{pe}$. Their definition of PMT quantum
efficiency is then $q_\text{pmt}=\Phi_\text{pe}/\Phi$ where $\Phi$ is
the incident photon flux from the absolutely calibrated beam.

We have reproduced this exact experimental setup and the beam geometry
in our PMT simulation (see Fig.~\ref{f:photonisExp}) in order to
``reverse engineer'' the PMT quantum efficiency $q_\text{pmt}$ into
the photocathode quantum efficiency $q_\text{pc}$. While using the
surface quantum efficiency expression from Eq.~\eqref{conversion} we
simulated a large number $N$ of the $\unit[375]{nm}$ photons in the
incident beam and recorded the number of released photoelectrons
$N_\text{pe}$. To reproduce the quoted PMT quantum efficiency
$q_\text{pmt}=0.26$ with our simulated fraction $N_\text{pe}/N$, the
conversion factor in Eq.~\eqref{conversion} had to be set to
$p_\text{conv}=0.405$.  Thus, in effect, at $\lambda=\unit[375]{nm}$
slightly more than 40\% of the absorbed photons in this geometry get
converted to photoelectrons.

\begin{figure}[!t]
\centering
\includegraphics[width=0.48\textwidth]{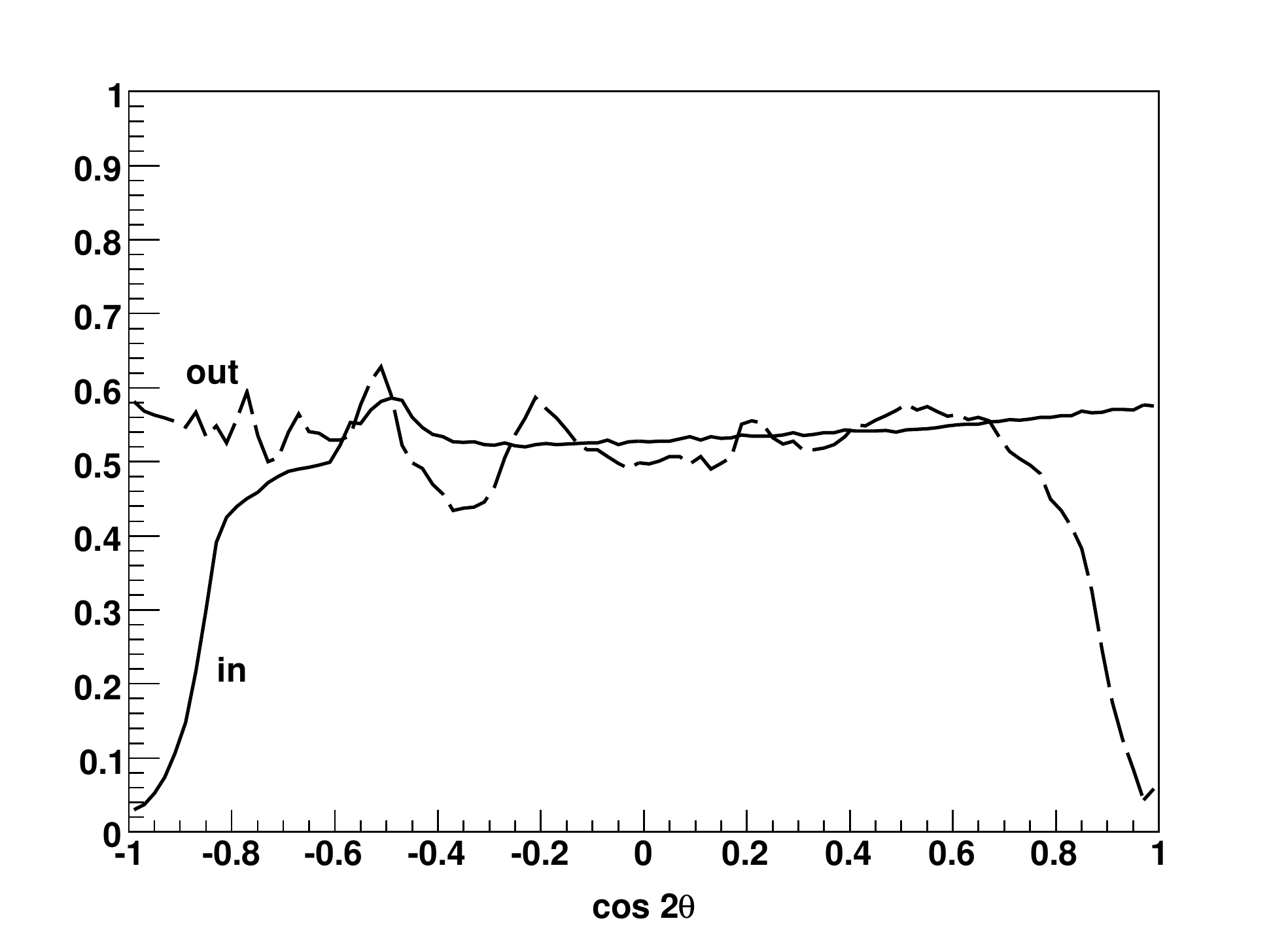}
\caption{Normalized distribution of photons arriving to the
  photocathode as a function of $\cos2\theta$ where $\theta$ is the
  incidence angle, for two populations, incoming (i.e.\ entering the
  PMT, full line) and outgoing (i.e.\ exiting the PMT, dashed line)
  photons. Due to the increased reflectivity of the water--glass
  interface a depletion from the isotropic distribution is observed
  for large incidence angles ($\cos(2\times90^\circ)=-1$) of incoming
  photons. For the outgoing photons the small incidence angles
  ($\cos(2\times0^\circ)=1$) are suppressed due to the reflection and
  absorption on the inner structure of the PMT. Note that the
  distribution for isotropic arrivals as a function of $\cos2\theta$
  is constant and equals to 1/2.}
\label{f:photonDistrib}
\end{figure}

\begin{figure}[!t]
\centering
\includegraphics[width=0.48\textwidth]{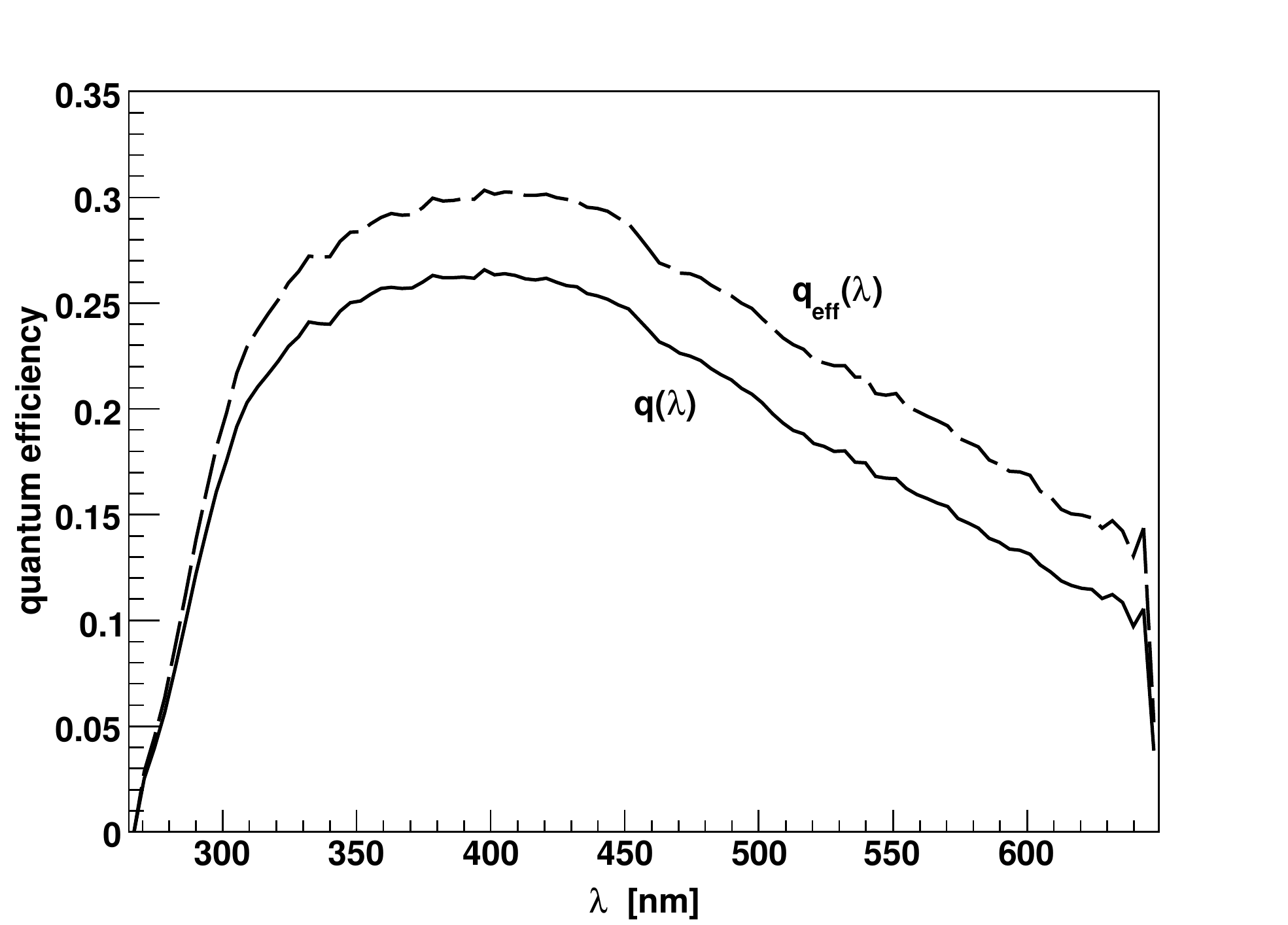}
\caption{Quantum efficiency $q$ (full line) defined as the probability 
for a photon to free an electron at the photocathode crossing. The 
dashed line corresponds to the effective quantum efficiency 
$q_\text{eff}$, i.e.\ the probability for an \emph{incoming} photon to 
eventually free an electron. The effective quantum efficiency is used in 
the implementation of the simplified PMT model.}
\label{f:QEW}
\end{figure}

\begin{figure}[t]
\centering
\includegraphics[width=0.48\textwidth]{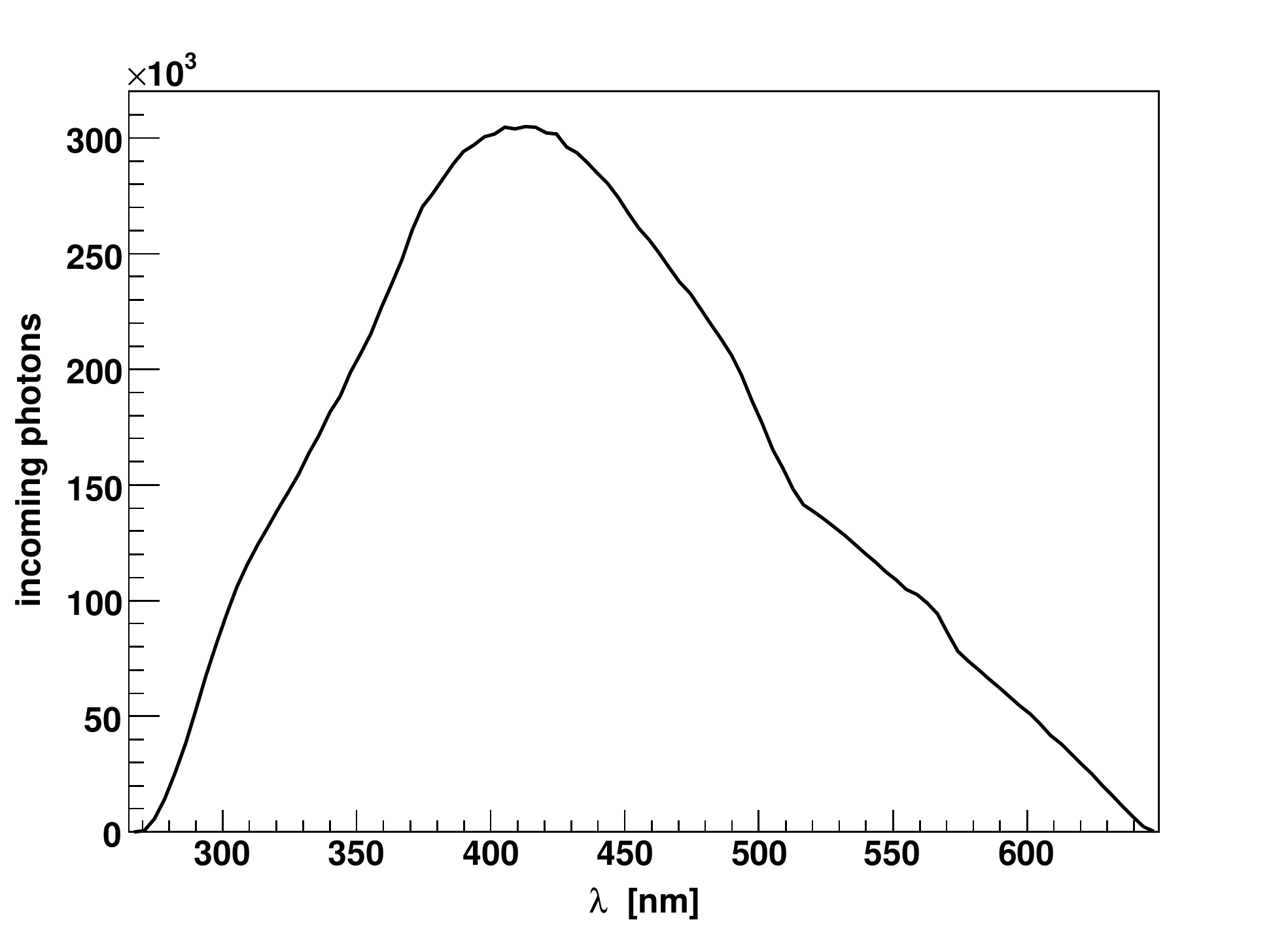}
\caption{Photon spectrum reaching the photocathode for 10\,000 injected 
vertical-centered muons. The main features of the curve are due to the 
shape of the Cherenkov emission spectrum, the wavelength dependency of 
the water absorption length (hump at $\sim\unit[420]{nm}$), the 
wavelength dependence of the reflectivity of the inner walls, and the 
PMT glass transparency (cut-off at wavelengths $<\unit[280]{nm}$).}
\label{f:photonSpectrum}
\end{figure}

\subsection{Quantum efficiency}

Using the obtained conversion factor $p_\text{conv}=0.405$, the PMT
simulation is run for a large number of injected vertical-centered
muons. Fig.~\ref{f:photonDistrib} shows the distribution of the 
incidence angle on the photocathode for photons entering (incoming) and 
exiting (outgoing) the PMT. On each crossing of the photocathode, 
absorbtance $A(\lambda,\theta)$ is obtained form the 
Eq.~\eqref{absorptance} and the photoelectron is released with 
probability $A(\lambda,\theta)\,p_\text{conv}$, as given by 
Eq.~\eqref{conversion}.  We can derive the overall, incidence-angle 
averaged quantum efficiency of our PMT simulation model by dividing the 
number of released photoelectrons $N_\text{pe}$ by the number of photons 
$N$ reaching the photocathode (from any side) for the different 
wavelengths,
\eq{
q(\lambda)=\frac{N_\text{pe}(\lambda)}{N(\lambda)}.
\label{qe}
}
The resulting quantum efficiency is shown in Fig.~\ref{f:QEW} (full
line). This quantum efficiency, obtained indirectly from the refractive
index parametrization in Fig.~\ref{f:IndexesW}, reproduces well the
known features of other experimental data \cite{motta} and the
specification of the PMT manufacturer \cite{photonis-catalogue}.
Furthermore, it gives the correct estimates of the average number of
photoelectrons released by the traversing muon as measured in
scintillator-triggered experiments with Auger SD stations
\cite{vem,allison}.

\section{Relevance to other astrophysics experiments}

There are several important points that can be extracted from the
previous sections. For optical interfaces, the \geant framework
implements only the familiar Fresnel equations of geometrical optics
and does not include the physical optics relevant for a thin-film
photocathode, as given by Eqs.~(\ref{matrix}--\ref{absorptance}). The
dependence of the photocathode absorption (and consequently of the
quantum efficiency) on the incidence angle is correctly described only
by the thin-film equations which have to be implemented independently
of \geant.  The absorption increases up to 50\% for incidence angles
larger than the angle of total internal reflection for the
glass--vacuum interface. No such increase is observed for the reverse,
vacuum--glass transition of the photons.  Furthermore, the quantum
efficiency obtained from the manufacturer's calibration setup already
contains a nontrivial fraction of the efficiency increase due to the
reflectivity of the inner structure of the PMT. The ``intrinsic''
quantum efficiency of the photocathode layer thus has to be
reverse-engineered by a procedure similar to the one described in
Section \ref{s:conversion-calibration}.  Clearly, both effects are
most relevant in cases of experiments with a distribution of incoming
photons substantially different from those characteristic of
manufacturer's experimental setup.

This may be of relevance to experiments employing similarly large PMTs 
in various Cherenkov detectors, like Amanda, IceCube, and IceTop 
\cite{amanda,icecube,icetop}, Kamiokande \cite{kamiokande}, NESTOR, 
NEMO, and Antares \cite{nestor,nemo,antares}, MiniBooNE 
\cite{miniboone}, Xenon \cite{xenon}, Baikal and Tunka 
\cite{baikal-tunka}, and northern Auger Observatory \cite{auger-north}, 
just to name a few.

\section{Simplified simulation model}

In a simulation, a vertical-centered muon injected into the water of the 
SD station has a $\unit[1.2]{m}$ tracklength and typically releases 
$\sim67\,500$ Cherenkov photons in the energy range between 1.5 and 
$\unit[5]{eV}$ (250 to $\unit[830]{nm}$ in wavelength; see 
Fig.~\ref{f:photonSpectrum} for spectrum).  Additional $\sim10\%$ of 
photons get produced by the secondary delta rays, in total giving 
$\sim73\,100$ photons per muon.  All these photons require a complete 
tracking inside the water container by \geant. The photons on their way 
undergo the absorption in the water medium, get reflected and absorbed 
by the container walls.  On average, only $\sim1340$ (1.8\%) of all 
photons eventually reach the PMT photocathode.  As described in the 
previous sections, the photon tracks are then simulated inside the PMT 
and the probability of a photoelectron release is evaluated.  A large 
fraction of produced photons will never reach the PMT and additionally a 
fraction of the incoming photons will be absorbed without producing any 
photoelectrons.  To avoid the time-consuming simulation of these photons 
that do not contribute to the PMT signal, a simplified simulation has 
been devised.  The goal is to estimate the effective probability for an 
emerging photon to contribute to the signal and eliminate idle photons 
from the simulation.  The corresponding \emph{statistical thinning} can 
thus be applied already at the time of their production.

\subsection{Effective quantum efficiency for incoming photons}

After several reflections on the walls, the swarm of photons behaves 
similarly to a photon gas. As can be already seen from the 
incidence-angle distribution of photons on the photocathode (see 
Fig.~\ref{f:photonDistrib}), the distribution of arrival directions is 
close to isotropic. Instead of the detailed simulation of photon tracks 
inside the advanced PMT model we can reduce the complexity of the 
procedure by introducing the simplified model of the PMT. The quantum 
efficiency in Eq.~\eqref{qe} stemming from Fresnel equations of the 
advanced PMT model is replaced with an \emph{effective} quantum 
efficiency $q_\text{eff}$ that describes the probability for an 
\emph{incoming} photon to produce a photoelectron,
\eq{
q_\text{eff}(\lambda) = 
  \frac{N_\text{pe}(\lambda)}{N_\text{in}(\lambda)}.
\label{qeff}
}
The $q_\text{eff}$ is obtained from the advanced model by normalizing 
the statistics of the released photoelectrons by the number of incoming 
photons. The actual wavelength dependence of $q_\text{eff}(\lambda)$ is 
shown in Fig.~\ref{f:QEW} (dashed line).

\begin{figure}[!t]
\centering
\includegraphics[width=0.48\textwidth]{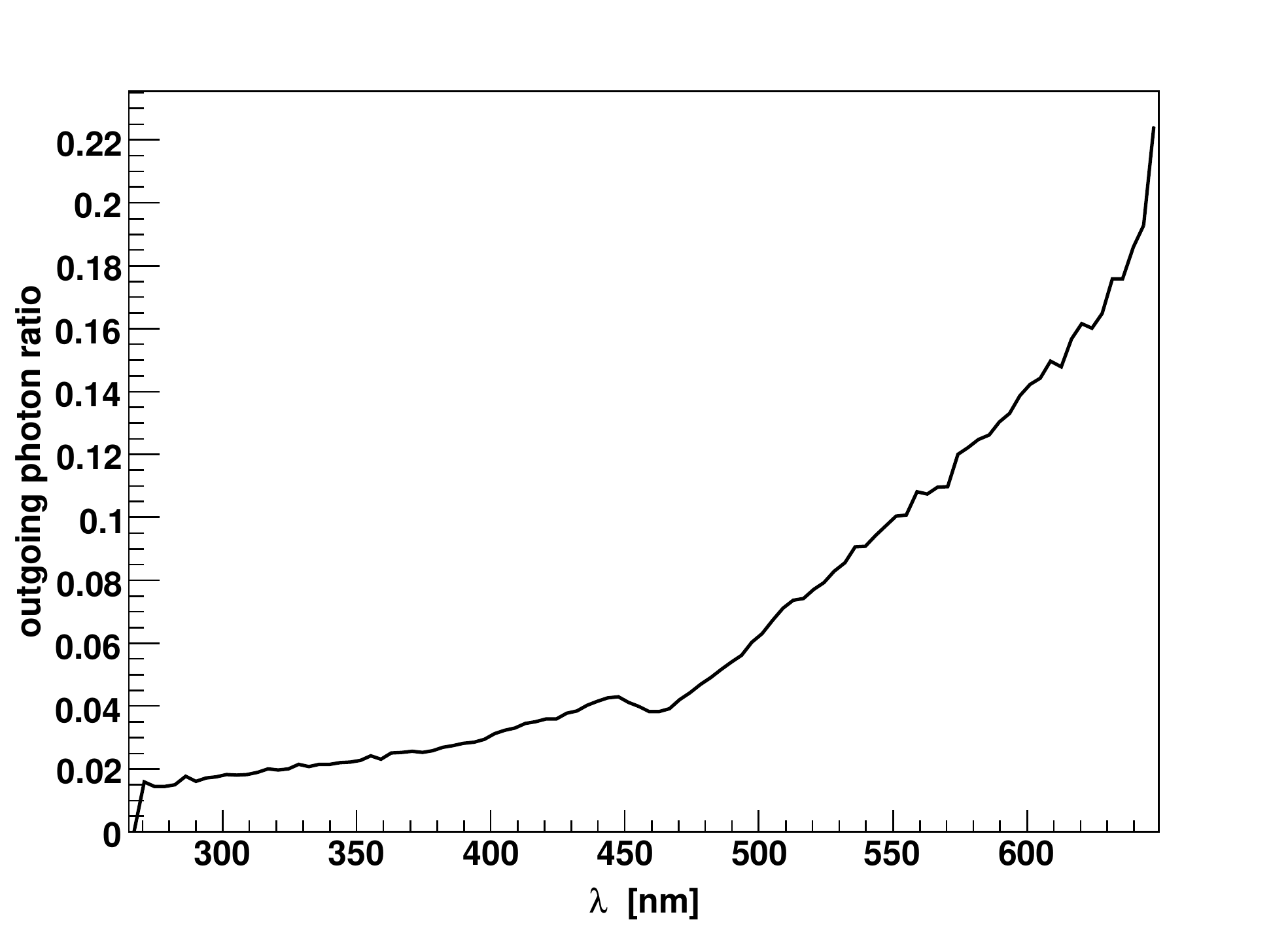}
\caption{Ratio of outgoing vs.\ incoming photons as counted at the PMT
photocathode.}
\label{f:out-over-in}
\end{figure}

In the advanced model only 4.5\% of the incoming photons return
to the water medium, i.e.\ only 0.08\% of all produced photons. In 
Fig.~\ref{f:out-over-in}, a wavelength-resolved fraction of outgoing 
vs.\ incoming photons is shown. The fraction is as low as 2\% for short 
wavelengths and does not exceed 22\% for long wavelengths.

\begin{figure}[!t]
\centering
\includegraphics[width=0.45\textwidth]{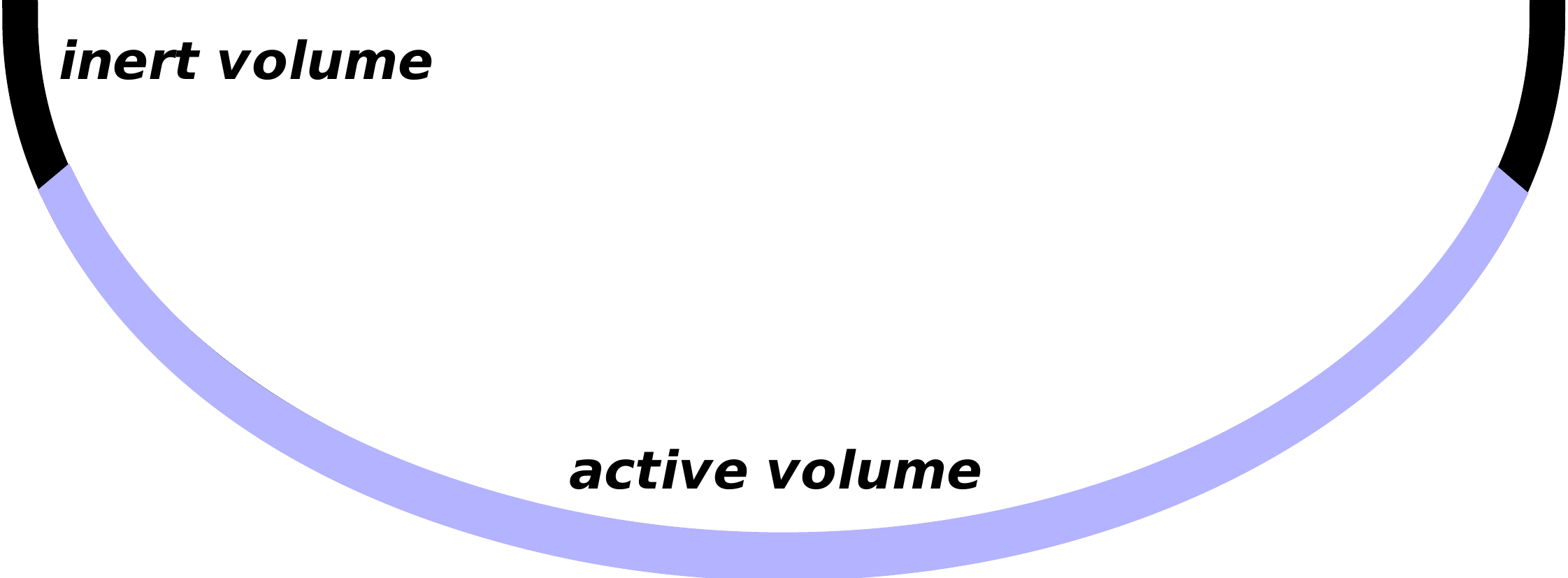}
\caption{Representation of the simplified model of PMT.}
\label{f:oldPMT}
\end{figure}

\begin{figure}[t]
\centering
\includegraphics[width=0.48\textwidth]{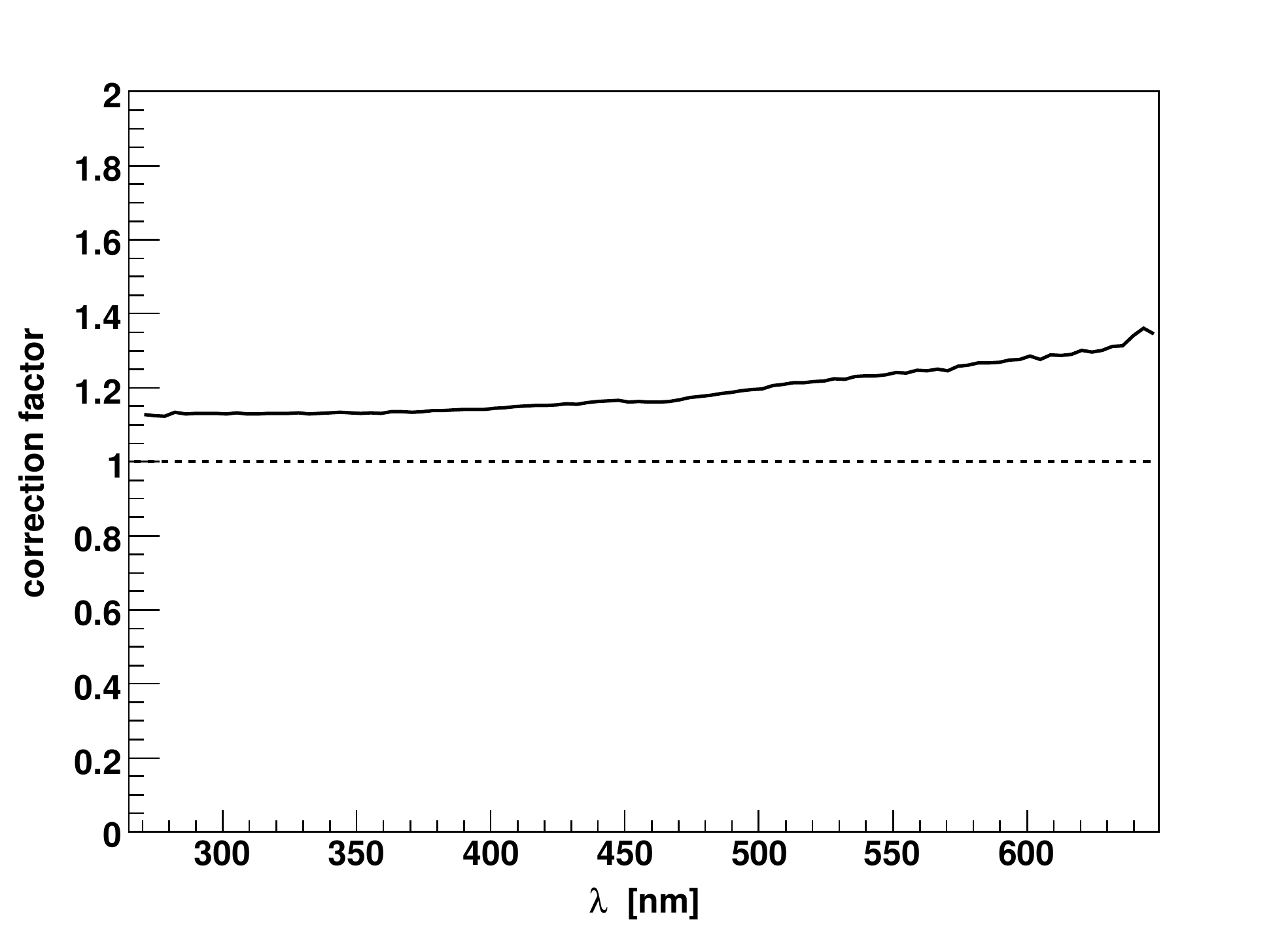}
\caption{Correction factor $q_\text{eff}(\lambda)/q(\lambda)$ as a 
function of the wavelength. For short wavelengths the correction is of 
the order of 13\% and goes up to 35\% for long wavelengths.}
\label{f:corrFactor}
\end{figure}

Based on these facts we have developed an implementation of the 
simplified PMT model that uses only a rudimentary geometry as shown in 
Fig.~\ref{f:oldPMT}. In this simplified model the photons are detected 
on contact with the outer surface and are consequentially removed from 
the simulation. No tracking of the photons on the inner structure of the 
PMT is performed. The chance of producing a photoelectron on the second 
crossing of the photocathode in the advanced model is accounted for in 
the simplified model by the usage of the effective quantum efficiency 
from Eq.~\eqref{qeff}. The relative ratio of the quantum efficiencies 
from Eqs.~\eqref{qeff} and \eqref{qe} is shown in 
Fig.~\ref{f:corrFactor}. The ratio is mostly close to $\sim10\%$ 
increase but then gradually climbs to $\sim35\%$ for long wavelengths.

\subsection{Statistical thinning}

Similar to the well established method in extensive air-shower
simulations \cite{mocca,corsika}, the concept of effective quantum
efficiency enables us to implement efficient statistical thinning of
the simulated photons.  At the moment of production, out of
$N_\text{pr}(\lambda)$ photons with wavelength $\lambda$ only
$q_\text{eff}(\lambda)N_\text{pr}(\lambda)$ photons with weight
$1/q_\text{eff}(\lambda)$ undergo further simulation.  As a result,
such thinned photons upon reaching the photocathode have to produce a
photoelectron with probability 1.  The average quantum efficiency (in
photon energy scale where Che\-ren\-kov spectrum is flat) is
$\bar{q}_\text{eff}=0.22$, resulting in a simulation speed-up for
almost a factor $\sim5$. If we take into account also the losses due
to the PMT collection efficiency, this number increases to $\sim6.5$.

\begin{figure}[t]
\centering
\includegraphics[width=0.48\textwidth]{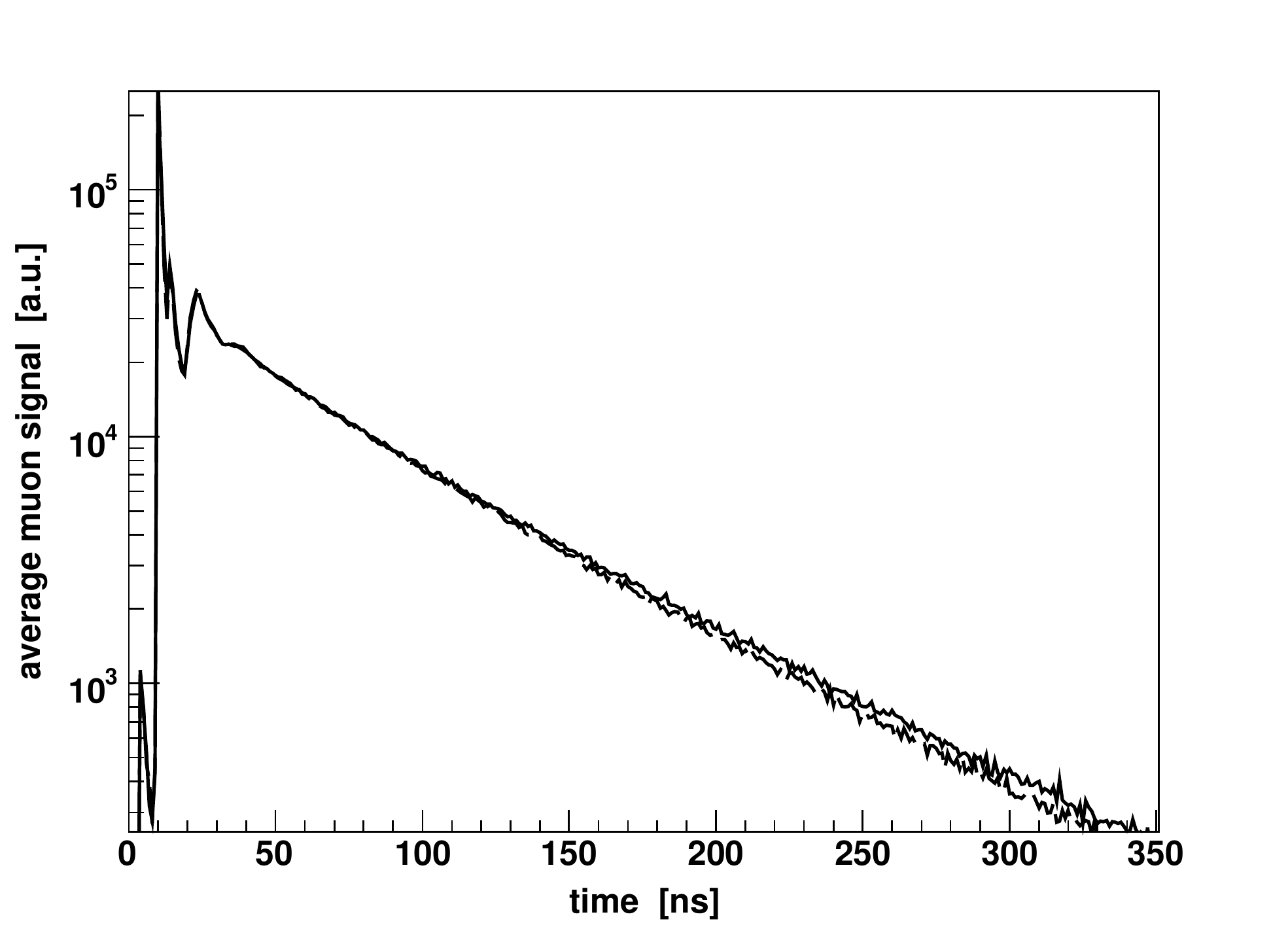}
\caption{Comparison of the time dependence of the muon signal in an SD
  station for the advanced PMT model (full line) and the simplified
  version (dashed line). The main feature is an approximate
  exponential time dependence with a decay time
  $\tau\approx\unit[65]{ns}$. The two models agree well with respect
  to the small details of the early signal and the latter decay
  part. A small discrepancy, limited to less than 5\%, is observed
  only in the late tail. Note that the oscillating structure for time
  $<\unit[30]{ns}$ is due to the nonuniform arrival of photons after
  the first few reflections when photons can not be treated yet as
  homogeneous and isotropic photon gas.}
\label{f:signal}
\end{figure}

In Fig.~\ref{f:signal} the results for the muon response in SD station 
is compared for both models. The main feature is an exponential decay 
but with a slightly changing exponent due to the wavelength dependence 
of the water absorption and wall reflectivity. The characteristic decay 
time of the muon signal is around
$\tau\approx\unit[65]{ns}$. Except in the tail of relatively large 
times, $t\gg\tau$, where the discrepancy is limited to within 5\%, the 
simplified model reproduces well the details of the muon signal from the 
advanced PMT model, especially the overall exponential decay and an 
oscillatory behavior for $t<\unit[20]{ns}$. It is worth mentioning here 
that the signal from the PMT in an SD station is sampled with a 
$\unit[40]{MHz}$ ($\unit[25]{ns}$) FADC, i.e.\ all these details will 
lie within one sampling bin.

\section{Summary}

We have implemented an advanced model for the simulation of the PMT 
response in a water-Cherenkov station of the surface detector. The model 
is based on the correct thin-film treatment of the photocathode optical 
processes with complex index of refraction. The model also includes 
tracking of the photons on the inner mirror surfaces and multiplier 
structure.

To relate the obtained quantum efficiency of the photocathode to quoted 
efficiency from the PMT specification we have in simulation reproduced 
the experimental set-up of the measurement done by the
manufacturer. This properly accounts for the increase in photon 
collecting due to the reflections on mirrored surfaces and inner 
structure of the PMT, and gives the average value of the conversion 
factor.

Since a relatively large number of photons is involved in the creation 
of the muon signal, certain details of the simulation, like the
dependence of the absorption on the incidence angles and the exact 
tracking of the photons on the inner structure, are washed out and can 
be treated in a phenomenological way.

To reduce the simulation time as well as the complexity of the problem 
we have implemented another, simplified model of the PMT.  Results for 
the average photoelectron release probabilities for different 
wavelengths in the advanced PMT model were in turn used to calibrate the 
effective quantum efficiency of the simplified model.  Results from both 
models agree well with the experimentally obtained number of produced 
photoelectrons per muon injection.  Finally, we have shown that the time 
dependence of the muon signal is almost not affected by the 
simplification in the simulation strategy.

\section*{Acknowledgments}

This work was supported by the Slovenian Research Agency and in part
by the Ad futura Programs of the Slovene Human Resources and
Scholarship Fund. Authors wish to thank Thomas Paul, Ralph Engel,
Slavica Kochovska, and the reviewer for useful suggestions, and the
many colleagues from the Pierre Auger collaboration that were involved
in SD detector calibration and simulation.

\end{document}